# Socratic Dialogs and Clicker use in an Upper-Division Mechanics Course

H. Vincent Kuo, Patrick B. Kohl, and Lincoln D. Carr

*Colorado School of Mines, Department of Physics, 1523 Illinois Street, Golden, CO 80401*

**Abstract.** The general problem of effectively using interactive engagement in non-introductory physics courses remains open. We present a three-year study comparing different approaches to lecturing in an intermediate mechanics course at the Colorado School of Mines. In the first year, the lectures were fairly traditional. In the second year the lectures were modified to include Socratic dialogs between the instructor and students. In the third year, the instructor used a personal response system and Peer Instruction-like pedagogy. All other course materials were nearly identical to an established traditional lecture course. We present results from a new instructor-constructed conceptual survey, exams, and course evaluations. We observe little change in student exam performance as lecture techniques varied, though students consistently stated clickers were "the best part of the course" from which they "learned the most." Indeed, when using clickers in this course, students were considerably more likely to become engaged than students in CSM introductory courses using the same methods.

## INTRODUCTION

The need for research-based curricular refinements in STEM disciplines is well-known, [1] but a complete restructuring of the physical instructional and learning environment requires significant resources and commitment by the instructor. And even though research has shown that such refinements [2-7] have significantly improved the quality of education at the introductory level with large enrollments, and thus can constitute a worthwhile investment, the benefits of restructuring an upper-division physics course are considered less-obvious. There have only been a limited number of cases where attempts have been made to reform physics major-only courses. [8-13]

The physics department at the Colorado School of Mines is large, regularly graduating approximately 55-60 majors per year, with a faculty that is strongly supportive of innovation in physics education (PER-based or otherwise). Perhaps not surprisingly, CSM has already seen some development of upper-division physics course reforms. [14,15] In this paper, we discuss reforms to an upper-division classical mechanics course, featuring three years' worth of data comparing two different lecture-based interactive engagement methods to a traditional lecture environment.

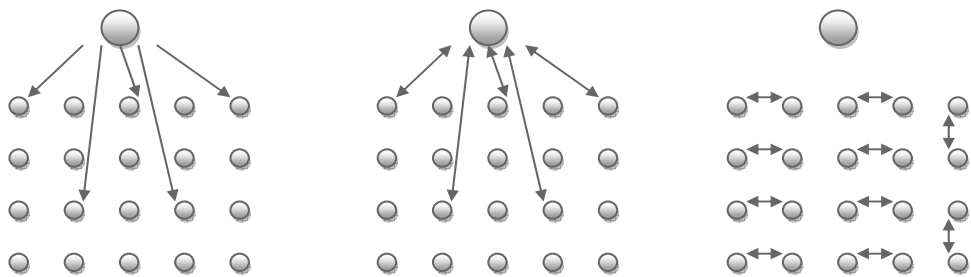

**Figure 1.** Graphical representation of the three different interaction styles. Arrows represent the interaction that is facilitated by each style. Left: Traditional Lecture (Fall 2008) where the instructor talks to the student body as a whole; the interaction is mostly uni-directional. Middle: Socratic Dialog (Fall 2009) where the instructor converses with individual students one to one; the interaction is bi-directional, one student at a time. Right: Peer Interaction (Fall 2010) where the instructor-designed clicker questions prompt students to talk to each other, and then with the instructor; the interaction is multi-directional and many interactions occur in parallel.

## COURSE AND METHODS

The course in question is the Intermediate Mechanics course taught in the junior year of the curriculum. The study took place during the fall of 2008, 2009, and 2010, and was taught by the same instructor in the same classroom with three different instructional styles (see Figure 1). The course met for 70 minutes three times a week, and was held in a traditional lecture hall environment. Although the instructor was aware of and enthusiastic about the results of recent physics education research, he is a traditional theorist and has no formal PER background

In the first year, the instructor taught in Traditional Lecture (TL) mode. This was also the first time that this instructor had taught this particular class. There were 63 students enrolled.

In the second year, the instructor committed to a pedagogy we refer to as Socratic Dialog (SD). In this mode, the instructor often intentionally engaged in one-on-one conversation with the students, soliciting questions and answers alike. This occurred regularly within each period, often following the introduction of a new concept or derivation. The instructor reported that he consciously attempted to engage every individual in the course several times during the semester by calling on them by name. There were 60 students enrolled.

In the third year, the instructor began using the personal response system favored in introductory physics courses at Mines (iClickers [16]) and adapted a Peer Instruction-like pedagogy. [6] In this approach, which we will call Peer Interaction (PI), the instructor posed the questions that he had used to start dialogs in the past in a multiple choice clicker format in an attempt to create an environment where the students are engaged more with each other than with the instructor. This occurred about as regularly as the Socratic dialogs in the previous year. There is an inherent external motivation in this peer-interaction mode that is different from past course offerings: a small portion of the course grade was tied to the correctness of the students' responses, with additional points given for merely participating. There were 55 students enrolled.

The content coverage in the course was similar from year to year, with some exceptions noted later. The major variation was in the lecture delivery method, with one additional confounding factor: The presence of a substitute lecturer for four of the sixteen weeks of the course. This substitute did not strictly follow the PI approach; we will return to this point.

There were three assessments that were used to characterize the effects of the different pedagogies: (1) an open-ended conceptual survey developed by the instructor of the course, given pre and post* (2) regular course exams; and (3) the university-administered course evaluations. Classroom videos were also taken 3 times during the PI semester to archive the level of student interaction and engagement.

## DATA

Table 1 shows data on the conceptual survey for the three semesters. All data are from the pre/post matched sample in each class, and the performance of the course is taken as a whole. The conceptual survey has ten questions and consists of standard topics in classical mechanics such as interpreting Lagrange's equations of motion, generalized coordinates, the Hamiltonian and total energy, and the inertia tensor. The gains shown are from a sub-set of the ten questions due to the fact that 4 of the topics were only marginally covered during the SD and PI semesters.

The pre-test was administered at the beginning of each semester, and was framed as a test "to see what you already know of the course material." The students were also told that they would receive credit for making a serious attempt and would receive a zero otherwise. The post-test was administered close to the end of the semester, and was presented as a way for the students to (1) prepare for the final, (2) see what they've learned from the course, and (3) compare to pre-test results for research purposes.

The post-test was given as an extra credit opportunity. Each administration of the survey took 50 minutes, and the identical survey was given as pre/post for each of the three years in the pilot study. The pre/post gain for the three different engagement styles are 0.74, 0.62, and 0.64 for TL (2008), SD (2009), and PI (2010), respectively. The differences between each of the treatment groups and the lecture group were statistically significant, but the difference between the SD and PI groups was not.

**Table 1**. Results from the instructor-constructed conceptual survey. The numbers here are from those students who have both pre- and post scores. The average percentage of the matched sample is reported for each year of the study, including the pretest, posttest, and the gain of the averages. Statistical significance test results are also included. These compared the pretest scores from group to group, the posttest scores, and the gains.

|  | Pretest Percentages | | | | Posttest Percentages | | | Gain |
|---|---|---|---|---|---|---|---|---|
|  | N | Ave. | StDev | StErr | Ave. | StDev | StErr |  |
| TL | 57 | 9.6 | 6.3 | 0.8 | 73.5 | 18.0 | 2.4 | 0.74 |
| SD | 54 | 11.8 | 5.7 | 0.8 | 65.2 | 17.3 | 2.4 | 0.62 |
| PI | 49 | 19.1 | 9.3 | 1.3 | 58.3 | 12.0 | 1.7 | 0.64 |
| t-test statistics | | | | | | | | |
| SD vs. TL | $p = 0.0526$, not significant | | | | $p = 0.011$, significant | | | $p < 0.0001$ |
| PI vs. TL | $p < 0.0001$, significant | | | | $p < 0.0001$, significant | | | $p < 0.0001$ |
| SD vs. PI | $p < 0.0001$, significant | | | | $p = 0.023$, significant | | | $p = 0.068$ |

Standard course exams provided another measure of student learning: two midterms and a cumulative final. The data presented here are the weighted average of all of the exams for each course (See Table 2). Once again, there were statistically significant differences between each of the treatment groups and the lecture group, but none between the treatment groups. The TL group performed better. Note that the exams given were not identical from year to year.

Students' perceptions of the course were reflected on a subset of the questions on the standard university-administered course evaluation. Scored on a 5-point Likert Scale, 5 being strong agreement with the question, the average scores on almost all of the relevant questions increased from TL to SD to PI:
  a) Teaching methods are effective: 2.83/2.96/3.07
  b) Created environment that fosters involvement: 3.04/3.20/3.54
  c) Facilitates student learning: 2.88/3.06/3.22
  d) Explains material clearly: 2.54/3.00/2.85
  e) Overall effectiveness: 2.90/3.00/3.24

Classroom interactions were videotaped several times during the course of the PI semester by one of the authors. The videos were reviewed to determine how many students were visibly interacting with one another, i.e., having a conversation. No effort was made to determine the content of the interactions. Over 90% of the students in the PI group regularly interacted with each other, with counts as high as 98% observed. Similar counts were done in CSM introductory physics courses using similar clicker pedagogies. There, 50-70% of the students visibly interact with one another, meaning that students in the upper division course were substantially more likely to participate in lecture-based interactive engagement techniques.

**Table 2.** Results from the exams. The numbers here are the weighted averages of the exams.

|  | Exams | | | |
|---|---|---|---|---|
|  | N | Ave. | StDev | StErr |
| TL | 57 | 71.8 | 10.1 | 1.4 |
| SD | 54 | 63.9 | 10.9 | 1.4 |
| PI | 48 | 65.9 | 9.8 | 1.4 |
|  | t-test statistics | | | |
| SD vs. TL | $p = 0.0001$, significant | | | |
| PI vs. TL | $p = 0.004$, significant | | | |
| SD vs. PI | $p = 0.33$, not significant | | | |

### DISCUSSION AND CONCLUSION

This paper presents results from a three-year long pilot study of the effects that different instructional pedagogies (Traditional Lecture, Socratic Dialog, and Peer Interaction) have on student learning in classical mechanics and their perceptions of the course.

We find that student learning as measured by a conceptual survey was higher in the Traditional Lecture semester. Exam data corroborated this finding. This is an interesting and surprising result; opposite the intentions of the instructor. However, there are notable confounding factors here. First, approximately one-quarter of the PI course was taught by a substitute postdoc who did not use the PI methodology (and was generally ill-received by the students), which almost certainly had an effect on the results for that semester. Second, as Table 1 shows, the pre-test scores varied significantly from trial to trial. We have not been able to identify the source of this variation, nor control for it. Comparing exams across semesters may also be a limitation of this study. Reliability and validity checks must be performed in the future to ensure comparability. We also do not have a satisfactory explanation or clear factors that we can attribute to the differences between the TL and SD treatments.

Next, students' self-reported preferences were for more interactive approaches, with higher scores for PI in every relevant course evaluation item except for "clarity of explanation." Unprompted free-response student comments specifically refer to the guest lecturer from the PI semester when criticizing the clarity of explanations, and also strongly praise the use of clickers in class.

Finally, student participation was very high in the interactive approaches. As seen from the video data, physics majors in this upper-division course participated almost universally in the interactive engagement portion of the course, as compared to lower-division courses wherein a number of students routinely click in and do not participate otherwise. This is somewhat surprising, as anecdotal faculty discussions sometimes raise the possibility that upper-division students don't need interactive engagement techniques to be successful, and may even react against them as being a tool for low-level classes.

These early results are intriguing, but call for additional methodological refinements and data. We will continue to explore how the different interactive engagement pedagogies affect student learning and perception starting in the fall of 2012. In particular, we will be repeating the PI implementation in the same Intermediate Mechanics course with more uniform instruction throughout and additional videotape data of student interaction. A more structured definition of what constitutes as "student engagement" in the video analysis will also be established.

## ACKNOWLEDGEMENTS

Special thanks to Tom Furtak and others in the CSM physics department for making educational reform a priority. Thanks also to N. Finkelstein for helpful discussions. This work was funded in part by NSF PHY #0903457.